# A Gentle Approach to Multi-Sensor Fusion Data Using Linear Kalman Filter

Parsa Veysi, Mohsen Adeli, Nayerosadat Peirov Naziri, Ehsan Adeli

*Abstract*—This research paper presents a detailed and approachable exploration of the Linear Kalman Filter (LKF) as a pivotal tool in the fusion of data originating from multiple sensors, providing a nuanced understanding of its capabilities in managing the intricacies of dynamic system estimations and predictions. The Kalman Filter, renowned for its recursive solution to the discrete-data linear filtering problem, is optimal for estimating states in dynamic systems, where it mitigates the impact of noise present in both measurements and the process. Our discourse is primarily concentrated on linear dynamic systems, due to the inherent assumptions of the LKF about system dynamics, measurement noise, and initial conditions. We meticulously elucidate the principles, assumptions, and operational mechanisms of the Linear Kalman Filter, and extensively detail its practical application in multi-sensor data fusion. This fusion is crucial, as it integrates diverse and heterogeneous sensory inputs to refine the accuracy and reliability of state estimations. To demonstrate the practical applicability and versatility of the LKF in real-world environments, this paper includes two physical examples where the Linear Kalman Filter significantly enhances the precision and stability in dynamic systems. These examples serve to illustrate the theoretical concepts discussed and provide tangible insights into the implementation of LKF in multi-sensor data fusion scenarios, emphasizing its paramount role in fields such as robotics, navigation, and signal processing. By combining a thorough exploration of the theoretical underpinnings of the Linear Kalman Filter with concrete practical applications and examples, this paper strives to offer comprehensive and accessible insights into the domain of multi-sensor data fusion. Our objective is to contribute meaningfully to the expanding corpus of knowledge in this vital area of research, fostering further innovations, advancements, and refinements in data fusion technologies, and facilitating their broader adoption across various scientific and industrial domain.

## I. Introduction

In a world inundated by a relentless surge of data, the synthesis of information from diverse and disparate sources is crucial. This is particularly true for dynamic systems, which are prevalent across a multitude of domains including, but not limited to, robotics, navigation, and signal processing. These systems, characterized by their continuous interaction and evolution within their environments, necessitate the employment of sophisticated methodologies capable of deciphering, managing, and interpreting the multifaceted data they produce [1]. One such methodology that stands out is the Linear Kalman Filter (LKF), a mathematical algorithm that has emerged as an invaluable asset in the field of data fusion, proficiently amalgamating data from multiple sensors and offering a nuanced understanding of its capabilities in managing the intricacies of dynamic system estimations and predictions [2]. The Linear Kalman Filter, with its capacity to attenuate the influence of noise inherent in measurements and processes, ensures the extraction of coherent and reliable information from environments saturated with uncertainty and noise. It offers recursive solutions to the discrete-data linear filtering problem, optimizing the estimation of states in dynamic systems.

This paper endeavors to delve into the depths of the Linear Kalman Filter, unraveling its principles, assumptions, operational mechanisms, and its practical implications in the realm of multi-sensor data fusion. The study is primarily concentrated on linear dynamic systems, a focus driven by the inherent assumptions of the LKF about system dynamics, measurement noise, and initial conditions. The fusion of diverse and heterogeneous sensory inputs is pivotal, serving to refine the accuracy and reliability of state estimations [3] and, consequently, enhancing the overall performance and reliability of dynamic systems. This research not only meticulously elucidates the theoretical underpinnings and operational intricacies of the Linear Kalman Filter but also shines a spotlight on its practical applications, providing tangible insights into its implementation in various multi-sensor data fusion scenarios. It aims to bridge the theoretical constructs with real-world applicability, demonstrated through illustrative physical examples, thereby offering a comprehensive perspective on the significant enhancement of precision and stability in dynamic systems facilitated by the LKF. By intertwining a profound exploration of the theoretical aspects of the Linear Kalman Filter with concrete practical applications and examples, this research aspires to present comprehensive and accessible insights into the domain of multi-sensor data fusion [4].

It is poised to make substantial contributions to the ever-expanding body of knowledge in this essential field of research, fostering further innovations, advancements, and refinements in data fusion technologies. Furthermore, it anticipates facilitating the broader adoption and implementation of these technologies across a spectrum of scientific and industrial domains, reflecting the transformative impact and paramount significance of the Linear Kalman Filter in the contemporary technological landscape. This introduction is a precursor to a more detailed exploration of the Linear Kalman Filter and sets the stage for a deeper, more nuanced understanding of its pivotal role in multi-sensor data fusion, opening the door to the intricate world of dynamic system estimations and predictions.

## II. BAYES' THEOREM

Bayes' Theorem, or Bayes' Rule, is a foundational principle [5] within probability theory and statistics, acting as a methodological compass for updating probabilities or beliefs based on new evidence. This theorem is at the heart of Bayesian inference, a paradigm allowing for the synthesis of prior knowledge in the examination and interpretation of uncertain phenomena, a capability crucial across various scientific and engineering domains including physics, computer science, economics, and biology. Bayes' Theorem is derived from the concept of conditional probability, a basic element of probability theory that quantifies the likelihood of one event occurring given the occurrence of another correlated event, symbolized as $P(A|B)$, representing the probability of event $A$ given the occurrence of event $B$.

$$P(A \mid B) = \frac{P(B \mid A) P(A)}{P(B)} \qquad (1)$$

In this equation, $P(A|B)$ is the posterior probability, representing the updated probability of event $A$ given the evidence $B$. $P(B|A)$ is the likelihood, indicating the probability of observing evidence $B$ assuming event $A$ occurs. $P(A)$ is the prior probability, depicting our initial belief or the intrinsic probability of event $A$ before new evidence is considered, and $P(B)$ is the evidence or marginal likelihood, illustrating the overall probability of observing evidence $B$. In the realm of Bayes' Theorem, each term holds significant meaning. The posterior probability is the refined probability of the hypothesis [6] after considering the new evidence. The likelihood is crucial for assessing how probable the observed evidence is, assuming the hypothesis is true. The prior probability is our initial degree of belief in the hypothesis before considering the new evidence, and the evidence is the observed data or information used to update the beliefs or probabilities, acting as the normalizing factor ensuring the sum of the posterior probabilities is one across all possible outcomes. Bayesian inference, grounded in Bayes' Theorem, has profound implications on the scientific method and our philosophical understanding of knowledge. It provides a coherent and logical framework for learning from data, allowing the integration of subjective beliefs with objective data to yield refined beliefs. This dynamic and adaptive process of belief updating is in harmony with the evolving nature of scientific knowledge [7], allowing for the constant refinement of theories and models as new data is acquired. The integration of prior knowledge is not only a methodological asset but also a philosophical stance acknowledging the cumulative and contextual nature of knowledge acquisition. It recognizes the substantial influence of pre-existing knowledge and beliefs on our understanding and interpretation and offers a formal mechanism for reconciling and integrating diverse sources of knowledge. The versatility and universal relevance of Bayes' Theorem have enabled its widespread application across diverse disciplines. In scientific research, it assists in the development and refinement of hypotheses and theories, allowing prior scientific knowledge to inform the interpretation of new experimental findings. In machine learning and artificial intelligence, it is employed for model learning and prediction, providing a principled approach to handling uncertainty and learning from data. In economics and finance, it is utilized for risk assessment and decision-making, allowing the incorporation of market insights and historical data to inform investment and policy decisions. Bayes' Theorem is not merely a mathematical equation; it is a foundational principle symbolizing the convergence of prior knowledge and new evidence in the quest for understanding and knowledge. It exemplifies the iterative and cumulative nature of scientific discovery, providing a structured and formalized methodology for the perpetual refinement of beliefs and theories considering new evidence. Its elegance, generality, and practicality have established it as an indispensable tool in the pursuit of knowledge, enabling advancements and discoveries across various domains of human inquiry [8][9].

## III. DYNAMIC SYSTEMS

In scientific and engineering domains, the study of dynamic systems—entities characterized by variable states over time—is indispensable. This paper investigates the multifaceted nature of dynamic systems, focusing on the interaction and evolution of constituent elements within a given system. Dynamic systems, ubiquitous across both natural phenomena and engineered structures, are delineated by the perpetual dynamics of forces, movements, and transformations.

This chapter elucidates the core principles and techniques requisite for the analysis of dynamic systems, contributing insights into their behavior, stability, and reaction to external inputs. It encompasses the examination of both linear and nonlinear systems, identifying the inherent patterns and intricacies within their frameworks. The primary methodology involves the application of mathematical models, predominantly differential equations, enabling the prediction and modification of system states, contingent on present conditions and governing laws. The exploration of varied applications and theoretical frameworks furnishes a comprehensive understanding of dynamic systems, empowering the modeling, analysis, and interpretation of intricate systems across diverse disciplines, from engineering to biology. This comprehension of dynamic systems lays the foundation for interpreting the relentless interplay and transformation characterizing the world, whether it is observed in the oscillations of a pendulum, the diffusion of heat, or the fluctuations in population dynamics. This research aims to augment the existing knowledge base, providing nuanced insights and applications pertinent to the field of dynamic systems.

### A. Differential Equation

A differential equation is a mathematical expression that interrelates one or more functions and their respective derivatives. Such equations are paramount in representing processes and phenomena observed across various fields [10], articulating the rate of change of physical quantities. Differential equations commonly appear in disciplines such as physics, biology, economics, and engineering, to model natural occurrences and analyze dynamic systems. Sir Isaac Newton, the inventor of calculus, utilized differential equations to formulate classical mechanics, describing the intricate relationships between position, velocity, and acceleration over time. For instance, if *x* denotes the position of a particle, its velocity *v* represents the rate at which the position alters with time.

$$v = \dot{x} = \frac{dx}{dt} \quad (2)$$

Similarly, acceleration *a* can be understood as the rate of change of velocity.

$$a = \dot{v} = \frac{d^2 x}{dt^2} \quad (3)$$

These equations, which express the dynamics of motion, are often referred to as the equations of motion. They serve as the foundation for understanding and interpreting the underlying principles governing the behavior and interaction of objects in the universe. They allow scientists and engineers to predict and analyze system behaviors, uncovering the inherent patterns and laws of the natural world [11][12].

*B. Differential Equation Types*

Differential equations are primarily classified into two categories: Ordinary Differential Equations (ODE) and Partial Differential Equations (PDE).

Ordinary Differential Equations are equations involving one independent variable and its derivatives. A substantial number of problems in physics are modeled using ODEs. For instance, equations of motion, which are functions of the derivatives of a single variable: position, fall under this category.

$$v = \frac{dx}{dt} \quad (4)$$

$$a = \frac{d^2 x}{dt^2} \quad (5)$$

In this context, time t is the independent variable, while *x* (position), *v* (velocity), and *a* (acceleration) are dependent variables since their values are contingent on *t*. ODEs are pivotal for data fusion and state estimation as they are instrumental in monitoring the evolution of variables over time, with time being the independent variable.

On the other hand, Partial Differential Equations encompass equations with multiple independent variables and their derivatives. The Laplace Equation is a classic example of a PDE, and it finds applications in various physics domains [13], including heat transfer and fluid dynamics.

$$\frac{\partial^2 u}{\partial t^2} = c^2 \frac{\partial^2 u}{\partial x^2} \quad (6)$$

PDEs are crucial in portraying complex systems [14] where multiple variables interact and influence each other, offering insight into multifaceted physical phenomena. This refined understanding and categorization of differential equations allow for a more nuanced approach to solving problems across different scientific disciplines, aiding in the development of advanced models and solutions.

*C. Dynamic Systems*

A system is construed as an assembly of interconnected components harmoniously interacting to function as a unified whole. When the attributes or properties inherent to the system exhibit temporal variations, it is characterized as a dynamic system. The evolution of the system over time is delineated as the process, and the underlying differential equations constituting the system are designated as the state equations of the dynamic system. Subsequently, the state variables of the system are interpreted as the dependent variables of the state equations [15][16].

Examine a system characterized by first-order differential equations that vary with time:

$$\begin{aligned} \dot{x}^1 &= f^1(t, x^1, x^2, ..., x^n, u^1, u^2, ..., u^m) \\ \dot{x}^2 &= f^2(t, x^1, x^2, ..., x^n, u^1, u^2, ..., u^m) \\ \dot{x}^3 &= f^3(t, x^1, x^2, ..., x^n, u^1, u^2, ..., u^m) \\ &\vdots \\ \dot{x}^n &= f^n(t, x^1, x^2, ..., x^n, u^1, u^2, ..., u^m) \end{aligned} \quad (7)$$

Given m states, the system is described by m state equations, with each equation being a function of time t, m states $x_i$, and $n_j$ inputs $u_j$. This system of equations can be succinctly expressed in vector form as:

$$\dot{x}(t) = f(t, x(t), y(t)) \quad (8)$$

Define the state vector *x(t)* and the input vector *u(t)* as follows:

$$x(t) = [x^1(t), x^2(t), ..., x^n(t)]^T \quad (9)$$

$$u(t) = [u^1(t), u^2(t), ..., u^m(t)]^T \quad (10)$$

Many real-world processes can be articulated through linear or nonlinear differential equations. Representing these processes in a state-space form enables the application of various mathematical methodologies to extract pertinent information and conduct comprehensive analyses of the system. Given the knowledge of the current state of the system and all current and subsequent inputs, it becomes feasible to prognosticate the values of future states and outputs of the system, along with additional insights. In essence, the state-space representation of dynamic systems opens avenues for enhanced understanding and in-depth analysis, allowing for the prediction and exploration of system behaviors, responses, and interactions over time, which is fundamental to advancing knowledge in fields reliant on dynamic systems modeling [17].

*D. Mathematical Model in Dynamic Systems*

Dynamic models predominantly employ two distinct representations of time, continuous-time, and discrete-time models, both yielding equivalent results but utilizing different mechanisms.

Continuous-time models represent time as a scalar real value t, providing a natural way of expressing time. In these models, the instantaneous rates, or derivatives, denoted by *x'(t)*, are calculated, forming the basis for differential equations representing the dynamic systems. To ascertain the state of the system at a future time point, it is imperative to integrate the differential equations, as illustrated below:

$$x(t) = \int_0^t f(t, x(t), u(t))\ dt \qquad (11)$$

Conversely, discrete-time systems segment the continuous time t into incremental time steps, representing the current time as an integer k multiplied by the time step, denoted as $\Delta t = k \cdot \Delta t$. In this representation, time, state, and input are all depicted at discrete intervals, where $t_k$, $x_k$, and $u_k$ symbolize the time, state, and input, respectively, at time step $k$. Unlike continuous-time models, discrete-time systems do not compute the derivative $x'(t)$ but rather determine the new state for the subsequent time step $k+1$ as demonstrated below:

$$x_{k+1} = f(t_k, x_k, u_k) \qquad (12)$$

While integration is not required in discrete-time systems, it is pertinent to note that these models can only be computed for discrete segments of time, contrasting with continuous-time models which permit calculations for any given time [18]. In conclusion, the choice between continuous and discrete representations hinges on the specific requirements and constraints of the modeling task, each providing unique insights and computational approaches to understanding dynamic systems.

*E. Mathematical Models in dynamic Systems*

The mathematical representation of dynamic systems in state space is a pivotal aspect of systems theory and control engineering, serving as a foundational framework for analyzing and understanding the inherent dynamics of a system. In this representation, dynamic systems can be articulated through two predominant temporal frameworks: continuous-time and discrete-time models. These frameworks offer distinct methodologies and insights, catering to various analytical needs and application domains [19].

**Continuous Linear Model**

The Continuous Linear Model is a specialized form of the continuous-time model, aptly represented as a linear operation [20] involving both the state vector, *x(t)*, and the input vector, *u(t)*. This model is a quintessential example of a system where the temporal evolution of the state is expressed as a linear function of the current state and input, providing a structured and analytical framework for studying dynamic systems.

The mathematical representation of the Continuous Linear Model is articulated as:

$$\dot{x}(t) = A(t)x(t) + B(t)u(t) \qquad (13)$$

Where:

- $\dot{x}(t)$ denotes the rate of change of the state vector,
- $A(t)$ is the state matrix representing the linear relationship between the state variables,
- $x(t)$ is the state vector,
- $B(t)$ is the input matrix defining the influence of the input variables on the state, and
- $u(t)$ is the input vector.

Expressed in expanded form, the equation becomes:

$$\begin{bmatrix} \dot{x}_1(t) \\ \dot{x}_2(t) \\ \vdots \\ \dot{x}_n(t) \end{bmatrix}_{k+1} = \begin{bmatrix} f_{11}(t) & f_{12}(t) & \cdots & f_{1n}(t) \\ f_{21}(t) & f_{22}(t) & \cdots & f_{2n}(t) \\ \vdots & \vdots & \ddots & \vdots \\ f_{n1}(t) & f_{n2}(t) & \cdots & f_{nn}(t) \end{bmatrix}_k \begin{bmatrix} x_1(t) \\ x_2(t) \\ \vdots \\ x_n(t) \end{bmatrix}_k$$
$$+ \begin{bmatrix} g_{11}(t) & g_{12}(t) & \cdots & g_{1n}(t) \\ g_{21}(t) & g_{22}(t) & \cdots & g_{2n}(t) \\ \vdots & \vdots & \ddots & \vdots \\ g_{n1}(t) & g_{n2}(t) & \cdots & g_{nn}(t) \end{bmatrix}_k \begin{bmatrix} u_1(t) \\ u_2(t) \\ \vdots \\ u_n(t) \end{bmatrix}_k \qquad (14)$$

**Characteristics of the Continuous Linear Model:**

- **Continuous-Time Nature**: This model calculates state rates, $\dot{x}$, representing the instantaneous rates of change of the state variables at any given continuous time point $t$.
- **Time-Varying Components**: The matrix $A(t)$ and $B(t)$ are functions of time, denoting the model's time-varying nature, where the linear relationship between variables may change over time.
- **Linear Operation**: The model is represented as a linear matric operation, showcasing the linear relationship between the state variables and the input variables within the system.

**Continuous Non-Linear Model**

The Continuous Non-Linear Model epitomizes the most versatile and generalized form of a continuous-time dynamic system. This model encompasses a broad spectrum of continuous systems, offering a universal representation that can accommodate any form of continuous dynamic system. It is characterized by its ability to represent complex, non-linear relationships between variables, making it an indispensable tool for studying a wide array of dynamic systems where linear approximations are inadequate.

The mathematical formulation of the Continuous Non-Linear Model is given by the following differential equation:

$$\dot{x}(t) = f(t, x(t), u(t)) \qquad (15)$$

Here:

- $\dot{x}(t)$ signifies the rate of change of the state vector,

- $f$ represents a general, possibly non-linear, function describing the relationships between time, state, and input,
- $x(t)$ is the continuous time variable,
- $x(t)$ denotes the state vector at time $t$,
- $u(t)$ is the input vector at time $t$.

**Characteristics of the Continuous Non-Linear Model:**

- **Continuous-Time Nature**: This model calculates the state rates, $\dot{x}$, elucidating the instantaneous rates of change of the state variables at any given continuous time point $t$.
- **Time-Varying Components**: The function $f$ is inherently dependent on time, indicating the time-varying nature of the model. This means the relationships and interactions within the system may evolve and change over time, allowing for the representation of dynamic systems with varying parameters and structures.
- **General Function Representation:** The model is encapsulated by a general function, $f$, representing potentially non-linear and complex relationships between the state and input variables. This generalized representation enables the model to depict a wide variety of dynamic behaviors and interactions, catering to the diverse needs of dynamic system analysis.

**Discrete Linear Model**

The Discrete Linear Model represents a specialized subclass of discrete-time models, characterized by its linear operations on the state and input vectors [21]. It provides a structured and concise representation of discrete dynamic systems, where the evolution of the system is described by linear relationships between the current state, the input, and the subsequent state.

The Discrete Linear Model is mathematically expressed as:

$$x_{k+1} = F_k x_k + G_k u_k \qquad (16)$$

Where:

- $x_{k+1}$ is the state vector at the next time step $k+1$,
- $F_k$ is the state transition matrix at time step $k$, defining the linear relationships between the state variables,
- $x_k$ denotes the state vector at the current time step $k$,
- $G_k$ represents the input matrix at time step $k$, characterizing the influence of the input variables on the state transition,
- $u_k$ is the input vector at time step $k$.

In expanded matrix form, the equation is represented as:

$$\begin{bmatrix} \dot{x}_1(t) \\ \dot{x}_2(t) \\ \vdots \\ \dot{x}_n(t) \end{bmatrix}_{k+1} = \begin{bmatrix} f_{11}(t) & f_{12}(t) & \cdots & f_{1n}(t) \\ f_{21}(t) & f_{22}(t) & \cdots & f_{2n}(t) \\ \vdots & \vdots & \ddots & \vdots \\ f_{n1}(t) & f_{n2}(t) & \cdots & f_{nn}(t) \end{bmatrix}_k \begin{bmatrix} x_1(t) \\ x_2(t) \\ \vdots \\ x_n(t) \end{bmatrix}_k \\ + \begin{bmatrix} g_{11}(t) & g_{12}(t) & \cdots & g_{1n}(t) \\ g_{21}(t) & g_{22}(t) & \cdots & g_{2n}(t) \\ \vdots & \vdots & \ddots & \vdots \\ g_{n1}(t) & g_{n2}(t) & \cdots & g_{nn}(t) \end{bmatrix}_k \begin{bmatrix} u_1(t) \\ u_2(t) \\ \vdots \\ u_n(t) \end{bmatrix}_k \qquad (17)$$

**Characteristics of the Discrete Linear Model:**

- **Discrete-Time Nature**: This model calculates the state, $x_{k+1}$, at the next time step, providing insights into the system's evolution in discrete intervals.
- **Time-Varying Components**: The matrices $F_k$ and $G_k$ are functions of the time step $k$, emphasizing the model's ability to represent systems with dynamics that may vary at each time step.
- **Linear Operation:** The model operates through linear matrix operations, reflecting the linear relationships and interactions within the system.

**Discrete Non-Linear Model**

The Discrete Non-Linear Model exemplifies the most encompassing and generalized representation within the realm of discrete-time dynamic systems. It offers a universal formulation capable of depicting any form of discrete system, making it an invaluable tool in the exploration and analysis of complex dynamic behaviors where linear models and continuous representations are not applicable [22].

The model is mathematically articulated as follows:

$$x_{k+1} = f(t_k, x_k, u_k) \qquad (18)$$

In this representation:

- $x_{k+1}$ denotes the state vector at the subsequent time step $k+1$,
- $f$ is a general function, potentially non-linear, illustrating the relationships and dependencies between the time, state, and input vectors,

- $t_k$ represents the discrete time step $k$,
- $x_k$ is the state vector at the current time step $k$,
- $u_k$ signifies the input vector at time step $k$.

Characteristics of the Discrete Non-Linear Model:

- **Discrete-Time Nature**: This model is inherent to discrete-time frameworks, calculating the states at discrete, sequential time steps, $x_{k+1}$, and allowing for the examination of system transitions and behaviors in quantized intervals.
- **Time-Varying Components**: The general function $f$ is a function of the time step $k$, indicating the model's capacity to represent time-varying dynamics, which is crucial for capturing evolving relationships and parameters within the system.
- **General Function Representation:** Represented by a general function, $f$, this model can embody complex, non-linear relationships between the state and input vectors, providing a versatile tool for the depiction of a wide spectrum of dynamic interactions and behaviors.

*F. Representation of Time-Invariant, Continuous-Time, Deterministic Linear Systems*

A time-invariant, continuous-time, deterministic linear system can generally be represented by the following differential equation:

$$\dot{x}(t) = Ax(t) + Bu(t) \tag{19}$$

Herein:

- $x$ is the state vector,
- $u$ is the control vector,
- $A$ and $B$ are the system and control matrices respectively, fundamentally defining the system's dynamic behavior.

The solution in continuous time is articulated as:

$$x(t) = e^{A(t-t_0)} x(t_0) + \int_{t_0}^{t} e^{A(t-\tau)} Bu(\tau) \, d\tau \tag{20}$$

Where:

- $x$ is the initial time, commonly set to zero,
- The matrix exponential term, often referred to as the state-transition matrix, delineates the state's evolution from an initial condition to a subsequent condition over a given period, in the absence of control involvement.

To transform the continuous-time system into its discrete counterpart, one introduces the discrete time relationship $t_k = t_{k-1} + \Delta t$ into the continuous-time solution, yielding:

$$x(t_k) = e^{A(t_k - t_{k-1})} x(t_{k-1}) + \int_{t_{k-1}}^{t_k} e^{A(t_k - \tau)} Bu(\tau) \, d\tau \tag{21}$$

Assuming a constant control input, u(t), over the interval ΔT and substituting v = τ − t k − 1, we obtain:

$$x(t_k) = e^{A\Delta T} x(t_{k-1}) + \int_{0}^{\Delta t} e^{A(\Delta t - v)} dv \cdot Bu(\tau) \, d\tau$$
$$= Fx(t_{k-1}) + Gu(t_{k-1}) \tag{22}$$

Here, the discrete-time matrices are defined as follows:

$$F = e^{A\Delta t} \tag{23}$$

$$G = F \int_{0}^{\Delta t} e^{-Av} dv \cdot B$$
$$= F(I - e^{-A\Delta t}) A^{-1} B, \; if \; A^{-1} \tag{24}$$

exists.

The matrix exponential, e At, can be computed or approximated through various methods, one of which is defined by the following series expansion:

$$e^{At} = \sum_{j=0}^{\infty} \frac{(At)^j}{j!}$$
$$= I + At + \frac{(At)^2}{2!} + \frac{(At)^3}{3!} + \ldots \tag{25}$$

A rudimentary approximation involves the truncation of higher-order terms, expressed as:

$$e^{At} \approx I + At \tag{26}$$

This approximation's validity is contingent upon the system matrix A; if higher-order terms significantly influence the outcome, alternative computational strategies must be employed [23][24].

IV. LEAST SQUARES ESTIMATION

In this section, we delve into the method of least squares estimation. This approach seeks to estimate a parameter by minimizing a cost function defined by the squared deviation from the observed data. Employing a quadratic cost function ensures the presence of a stationary point on the cost surface. This stationary point inherently represents the solution that minimizes the error, thereby providing the optimal estimate.

*A. Estimation of a Constant Scalar: A Mathematical Perspective*

In this section, we focus on the methodology for estimating a constant scalar from a sequence of noisy observations of that scalar. Consider the scenario where a temperature sensor, albeit of low precision, is attached to an engine, and our objective is to determine the engine's temperature. Due to the sensor's limited quality, each reading is prone to noise. Consequently, to improve the accuracy of our estimation, multiple readings are taken.

Formally, let $y_i$ represent a series of measurements, $i \in \{1,\ldots,k\}$, of an unknown scalar $x$ (in this context, the engine temperature). Each measurement is perturbed by an independent, white, zero-mean noise $v_i$, such that $E(V) = 0$ and $E(v_i v_j^T) = 0$, where $v_i \in V$. Thus, each observation can be modeled as:

$$y_i = x + v_i$$

To deduce the true value of $x$ from the set of observations, one intuitive approach is to compute the average of all readings:

$$\bar{y} = \frac{1}{k} \sum_{i=1}^{k} y_i \tag{27}$$

$$\bar{y} = \frac{1}{k} \sum_{i=1}^{k} (x + v_i) \tag{28}$$

$$\bar{y} = x + \bar{v} \tag{29}$$

Given the noise's distribution, we understand that its mean (or expected value) is zero:

$$\bar{v} = E(V) = 0 \tag{30}$$

Thus, our optimal estimation of $x$, denoted as $\hat{x}$, is derived from the mean of all the observations:

$$\hat{x} = \hat{y} \tag{31}$$

This methodology is predicated on the notion that, with a sufficiently large number of readings, the cumulative effect of the noise approaches a zero mean. The requisite number of measurements is contingent upon the desired estimation precision and the noise's magnitude [25][26].

### B. Linear Least Squares: Vector Estimation

Place Moving from the estimation of a constant scalar discussed previously, we now explore the estimation of a constant vector. Consider a noisy measurement $y_i$, which can now be expressed as a linear combination of the elements of the vector $x$ we aim to estimate:

$$y_1 = H_{11} x_1 + \ldots + H_{1n} x_n + v_1$$
$$\vdots \tag{32}$$
$$y_k = H_{k1} x_1 + \ldots + H_{kn} x_n + v_k$$

Our target is an n-dimensional vector, and the system of equations can be concisely represented in matrix form:

$$\begin{bmatrix} y_1 \\ y_2 \\ \vdots \\ y_k \end{bmatrix} = \begin{bmatrix} H_{11} & H_{12} & \cdots & H_{1n} \\ H_{21} & H_{22} & \cdots & H_{2n} \\ \vdots & \vdots & \ddots & \vdots \\ H_{k1} & H_{k2} & \cdots & H_{kn} \end{bmatrix} \begin{bmatrix} x_1 \\ x_2 \\ \vdots \\ x_n \end{bmatrix} + \begin{bmatrix} v_1 \\ v_2 \\ \vdots \\ v_k \end{bmatrix} \tag{33}$$

This can be succinctly written as:

$$y = Hx + v \tag{34}$$

Our objective is to determine x. While the noise vector v remains unknown, we aim to derive the best possible estimate $\hat{x}$. The error residual $\epsilon$ or the discrepancy between the measurements y and the estimated vector H x ^ is given by:

$$\varepsilon = y - H\hat{x} \tag{35}$$

Our goal is to minimize the magnitude of $\epsilon$. To achieve this, we define a cost function $J$ related to the error residual. The function $J$ is the sum of squared errors, defined as:

$$J = \varepsilon^T \varepsilon \tag{36}$$

$$J = (y - H\hat{x})^T (y - H\hat{x}) \tag{37}$$

$$J = y^T y - \hat{x}^T H^T y - y^T H\hat{x} + \hat{x}^T H^T H\hat{x} \tag{38}$$

To determine the minimum of $J$, we differentiate it with respect to $\hat{x}$ and equate the result to zero. This helps identify the stationary points:

$$\frac{\partial J}{\partial \hat{x}} = -2H^T y + 2H^T H\hat{x} = 0$$

Solving $x$, we obtain:

$$\hat{x} = (H^T H)^{-1} H^T y \tag{40}$$

This formula represents the least squares solution, minimizing the sum of squared errors. For solvability, matrix $H$ must be of full rank, and $(H^T H)$ must be invertible. Additionally, the number of measurements $k$ should exceed the number of elements n in $\hat{x}$. Revisiting the engine temperature scenario, let's determine the relationship between engine temperature and its speed, denoted as RPMs (revolutions per minute). Given engine temperature measurements $y_i$ at various RPM speeds $r_i$, we aim to derive a linear relationship of the form $y = ax + b$, which can be represented as:

$$y_i = x_1 r_i + x_2 + v_i \tag{41}$$

Using the least squares solution, the relationship in represented as:

$$\begin{bmatrix} y_1 \\ y_2 \\ \vdots \\ y_k \end{bmatrix} = \begin{bmatrix} r_1 & 1 \\ r_2 & 1 \\ \vdots & \vdots \\ r_k & 1 \end{bmatrix} \begin{bmatrix} x_1 \\ x_2 \end{bmatrix} + \begin{bmatrix} v_1 \\ v_2 \\ \vdots \\ v_k \end{bmatrix} \tag{42}$$

Here, the estimated vector $\hat{x}$ represents the coefficients of the best-fit line.

### C. Weighted Least Squares: Incorporating Noise Information

In the preceding discussions concerning least squares estimation, the noise components $v_i$ were not explicitly incorporated into the solutions. In scenarios where we possess specific information regarding the noise, particularly about the accuracy of each measurement or the relative reliability of some measurements over others, it becomes pertinent to integrate this knowledge into the estimation framework. Leveraging this information can potentially refine the accuracy of our solution.

To elaborate further, consider a scenario where the vector $x$ to be estimated remains a constant $n$-dimensional vector, and $y$ is a $k$-dimensional noisy measurement vector that represents a linear combination of $x$ via the model matrix $H$. Each element in this matrix incorporates additive noise components $v_i$ with a variance $\sigma_i^2$. This scenario can be mathematically depicted as:

$$\begin{bmatrix} y_1 \\ y_2 \\ \vdots \\ y_k \end{bmatrix} = \begin{bmatrix} H_{11} & H_{12} & \cdots & H_{1n} \\ H_{21} & H_{22} & \cdots & H_{2n} \\ \vdots & \vdots & \ddots & \vdots \\ H_{k1} & H_{k2} & \cdots & H_{kn} \end{bmatrix} \begin{bmatrix} x_1 \\ x_2 \\ \vdots \\ x_n \end{bmatrix} + \begin{bmatrix} v_1 \\ v_2 \\ \vdots \\ v_k \end{bmatrix} \quad (43)$$

where

$$E(v_i^2) = \sigma_i^2 \quad (i = 1, \ldots, k) \quad (44)$$

and

$$E(vv^T) = R = diag(\sigma_1^2, \sigma_2^2, \ldots, \sigma_k^2) \quad (45)$$

Our objective is to minimize the cost function $J$ concerning $\hat{x}$ such that:

$$J = \frac{\varepsilon_1^2}{\sigma_1^2} + \frac{\varepsilon_2^2}{\sigma_2^2} + \ldots + \frac{\varepsilon_k^2}{\sigma_k^2} = \varepsilon^T R^{-1} \varepsilon \quad (46)$$

This revised cost function optimizes the sum of the squared errors, each weighted by its corresponding noise variance (hence "weighted least squares"). A greater variance implies a lesser weight for that measurement. It's the relative weights that are crucial. If all weights were uniform, the solution would coincide with the standard least squares solution [27].

Differentiating the cost function and identifying the point of minimal value by equating the derivative to zero, we get:

$$J = \varepsilon^T R^{-1} \varepsilon = (y - H\hat{x})^T R^{-1}(y - H\hat{x}) \quad (47)$$

Upon differentiation,

$$\frac{\partial J}{\partial \hat{x}} = -2H^T R^{-1} y + 2H^T R^{-1} H\hat{x} \quad (48)$$

leading to

$$H^T R^{-1} y = H^T R^{-1} H\hat{x} \quad (49)$$

Yielding

$$\hat{x} = (H^T R^{-1} H)^{-1} H^T R^{-1} y \quad (50)$$

The uncertainty in the estimates can be assessed using uncertainty error propagation. If a linear relationship $f = Ax$ exists, the uncertainty covariance matrix of $x$, $\Delta x$, can be transformed into the uncertainty covariance matrix of $f$, $\Delta f$, through $\Delta f = A\Delta x A^T$. Applying this to the weighted least squares solution:

$$\Delta \hat{x} = (H^T R^{-1} H)^{-1} H^T R^{-1} \Delta y (H^T R^{-1} H)^{-1} H^T R^{-1} \quad (51)$$

Assuming the uncertainty is modelled accurately such that $\Delta y = R$, the equation becomes:

$$\Delta \hat{x} = (H^T R^{-1} H)^{-1} \quad (52)$$

### D. Recursive Least Squares: Dynamic Estimation Incorporating Sequential Measurements

In the context of least squares estimation, a typical approach involves constructing a model matrix $H$ from various measurements and subsequently solving a matrix equation [28]. This strategy assumes that all measurements are simultaneously accessible during the estimation. However, in real-time scenarios, measurements might be obtained sequentially over time. Storing all past measurements and augmenting the $H$ matrix can be cumbersome and computationally inefficient. An alternative is to recursively update the estimate with each new measurement, obviating the need to retain a history of prior measurements and relying only on the most recent estimate [29][30].

Consider $\hat{x}_k$ as the estimated constant n-dimensional vector which integrates all measurement data up to the $k$-th observation. Let $y_k$ signify the $k$-th noisy measurement vector, representing a linear combination of $x$ through the model matrix $H_k$ and perturbed by random additive noise $v_k$. A linear recursive estimator can be represented as:

$$y_k = H_k x + v_k \quad (53)$$

$$\hat{x}_k = \hat{x}_{k-1} + K_k (y_k - H_k \hat{x}_{k-1}) \quad (54)$$

Here, the updated estimate $\hat{x}_k$ is derived from the preceding estimate $\hat{x}_{k-1}$ and the latest measurement $y_k$. The deviation from the prior estimate is dictated by the discrepancy between the current measurement $y_k$ and the estimated measurement derived from $H_k \hat{x}_{k-1}$. This discrepancy is scaled by a gain matrix $K_k$, optimized to minimize a specific least squares cost function.

Examining the estimation error, $\epsilon_k = x - \hat{x}_k$, we can express it recursively as:

$$\varepsilon_k = (I - K_k H_k)\varepsilon_{k-1} - K_k v_k \quad (55)$$

The estimation error covariance matrix $P_k$, reflecting the estimation error, is defined as:

$$P_k = \varepsilon_k \varepsilon_k^T \quad (56)$$

By expanding $P_k$ in a recursive form and utilizing properties of the noise, the covariance matrix evolves as:

$$P_k = (I - K_k H_k) P_{k-1} (I - K_k H_k)^T + K_k R_k K_k^T \quad (57)$$

To minimize the cumulative estimation errors, we define a cost function $J_k$:

$$J_k = \sum_{i=1}^{k}(x_i - \hat{x}_i)^2 = \varepsilon_k^T \varepsilon_k \qquad (58)$$

Minimizing $J_k$ is equivalent to minimizing the trace of the estimation error covariance matrix $P_k$.

The optimization goal becomes determining the gain matrix $K_k$ that minimizes the estimation error, steering $\hat{x}_k$ towards $x$. After differentiating the cost function and equating the result to zero, we find:

$$K_k = P_{k-1} H_k^T (H_k P_{k-1} H_k^T + R_k)^{-1} \qquad (59)$$

For implementing the recursive least squares, initiate the estimator with:

$$\hat{x}_0 = E(x) \qquad (60)$$

$$P_0 = E[(x - \hat{x}_0)(x - \hat{x}_0)^T] \qquad (61)$$

Here, $\hat{x}_0$ represents the initial guess or best estimate, and $P_0$ denotes the uncertainty of this estimate. With each new measurement $y_k$, the solution updates as:

$$K_k = P_{k-1} H_k^T (H_k P_{k-1} H_k^T + R_k)^{-1} \qquad (62)$$

$$\hat{x}_k = \hat{x}_{k-1} + K_k (y_k - H_k \hat{x}_{k-1}) \qquad (63)$$

$$P_k = (I - K_k H_k) P_{k-1} (I - K_k H_k)^T + K_k R_k K_k^T \qquad (64)$$

## V. KALMAN FILTER

The nomenclature "Kalman Filter" is derived from its innovator, Rudolf Kalman, born in 1930 in Budapest [31]. He migrated to the U.S. during World War II and pursued his higher education at MIT, leading to his role as a professor at Columbia University. The first practical implementation of the Kalman Filter was by NASA's Ames Research Center and was subsequently utilized for the Apollo spacecraft mission to estimate trajectory and control.

### A. The Kalman Filter: Simplifying Data Fusion in Linear Dynamic Systems

The Kalman Filter is recognized as a pivotal method for transforming the intricate and potentially computationally burdensome procedure of data fusion into a more manageable problem [32]. This simplification is achieved by incorporating several assumptions regarding system dynamics, estimations, states, and the properties of errors and noise. From a mathematical perspective, the Kalman Filter provides a solution to a linear quadratic estimation problem. Essentially, it formulates estimates and processes the instantaneous data of a linear dynamic system disturbed by random white noise by utilizing a sequence of measurements. These measurements are linearly related to the state but are also distorted by white noise.

$$\hat{x} = E = (x \mid z_1, z_2, \cdots, z_k) \qquad (65)$$

$$\hat{x} = \min J \qquad (66)$$

$$\tilde{x} = x - \hat{x} \qquad (67)$$

$$P = [(x - \hat{x})(x - \hat{x})^T] \qquad (68)$$

$$J = [(x - \hat{x})^T (x - \hat{x})] = Tr(P) \qquad (69)$$

The objective is to compute the estimated state that is closest to the true, albeit unknown, state using a series of noisy measurements. The estimated state solution is statistically optimal concerning the quadratic function of the estimation error, minimizing the squared mean error analogous to the Least Squares method. The pursued cost function for minimization mirrors the one employed in Recursive Least Squares, focusing on reducing the state estimation error, the discrepancy between the true state and the estimate. The covariance matrix of the state estimation error and the cost function associated with the squared mean of the state estimation error are articulated correspondingly, and minimizing the trace of the covariance matrix concurrently minimizes the cost function. This methodology bears a significant resemblance to the Recursive Least Squares process. Further elaboration on the Kalman Filter will be provided subsequently. While probability has not been explicitly mentioned, the representation of the state as a probability distribution has been previously discussed. By making specific assumptions about noise and system dynamics, the probabilistic problem can be transformed into an optimization problem and solved accordingly. The Kalman Filter can be viewed from optimal, geometric, or probabilistic perspectives, but invariably leads to identical equation solutions, which is remarkable. It is hailed as one of the most significant discoveries in the fields of estimation and data fusion, enabling advancements in numerous complex dynamic systems such as those in guidance, navigation, and control of various vehicles and used extensively in domains like robotics, manufacturing, signal processing, economics, and finance [33][34].

### B. Types of Kalman Filters

Several variants of the Kalman Filter merit discussion, each distinguished by its operational spectrum and application. Primarily, the Linear Kalman Filter is addressed in this discourse, serving as the foundation for understanding the subsequent types. Subsequently, the Extended Kalman Filter is introduced, capable of operating on a broader range of non-linear systems compared to its linear counterpart. Further advancement is seen in the Unscented Kalman Filter, which exhibits superior performance on non-linear systems beyond the capabilities of the Extended Kalman Filter. The Linear Kalman Filter operates under the assumption of linear system dynamics, which can be either discrete or continuous. Its methodology employs linear covariance prediction and update equations during the filtering process. However, due to its inherent assumption of linear system dynamics, the Linear Kalman Filter is incompatible with non-linear systems.

Moving forward, the Extended Kalman Filter is introduced to adapt to non-linear system dynamics, marking a deviation from the Linear Kalman Filter. This adaptation can be applied to both discrete and continuous system dynamics. The Extended Kalman Filter utilizes linear covariance prediction and update equations, approximating linear relationships from non-linear system dynamics, proving effective especially with smooth non-linear systems where the estimated state closely mirrors the true state. Nonetheless, its limitation lies in its inability to apply to the entire spectrum of non-linear systems. It operates on the premise that the system can be linearized around the current state estimate if this estimate closely aligns with the true state. Lastly, the Unscented Kalman Filter is designed to assume non-linear, discrete, or continuous system dynamics, serving as an analogy to the Extended Kalman Filter. It integrates system nonlinearities to compute prediction and update equations, thus offering a more accurate approximation of the system dynamics. The Unscented Kalman Filter maintains approximation by calculating a linear approximation of the system variances, albeit at a different phase of the process. Its significant advantage is its efficiency in handling more pronounced non-linear systems due to its superior approximation capabilities. It acts as a higher-order approximation compared to the first-order approximation in system dynamics, seen in the Extended Kalman Filter. Each variant of the Kalman Filter encapsulates unique methodologies and applications, with advancements addressing the limitations and expanding the operational scope of their predecessors. The exploration of these filters provides insight into their distinctive capabilities and constraints, contributing to the optimized application of data fusion techniques across diverse systems and domains [35][36][37].

*C. Working Principle of the Kalman Filter*

The Kalman Filter is predicated on the assumption that all probability distributions associated with state and measurement can be aptly represented as Gaussian distributions [38]. This assumption, albeit an approximation, greatly simplifies Bayesian calculations and has proven effective in practical applications, positioning the Kalman Filter as a premier method for data fusion within various industries.

$$x_k = F_{k-1} x_{k-1} + G_{k-1} u_{k-1} + L_{k-1} w_{k-1} \qquad (70)$$

$$z_k = H_k x_k + M_k v_k \qquad (71)$$

The methodology of the Kalman Filter involves the propagation of the mean and covariance of Gaussian distributions [39] for the estimated state through time, utilizing a linear process model. This is known as the prediction or process update step, where estimates are advanced in time whenever linear combination elements of the state are available. Subsequently, the Kalman Filter refines the estimated state mean and covariance based on measurements made and the covariance of the Gaussian distribution, referred to as the update or correction step. These steps form a continuous prediction-correction cycle, recursively run as time progresses, with estimates being continually refined and propagated to the current time using available measurement information. Given the inherent discreteness of any implementation of the linear Kalman Filter, our initial focus is on the discrete Kalman Filter, which operates in discrete time blocks for systems that are inherently smooth and continuous. In such discrete systems, time progresses in blocks or steps, causing corresponding alterations in the system state. When the responses of continuous and discrete systems are overlaid, congruence is observed at discrete key points in time. For linear, discrete-time systems, the process model illustrates how the system evolves, functioning based on the current state, control inputs, and noise inputs, while the measurement model is derived as a linear combination of the current state and some form of random noise.

$$\hat{x}_k^+ = E(x_k \mid z_1, z_2, \cdots, z_k) \qquad (72)$$

$$\hat{x}_k^- = E(x_k \mid z_1, z_2, \cdots, z_{k-1}) \qquad (73)$$

$$x_0^+ = E(x_0) \qquad (74)$$

In employing the Kalman Filter, several critical assumptions are made regarding the noise within the system. It is presumed to follow a Gaussian distribution with zero mean and a specified covariance matrix, and the noise variables are considered uncorrelated over time and independent between the process and measurement noise. The principal aim is to accurately estimate the state of the system, leveraging knowledge of the dynamics and the availability of noise measurements, to attain a posterior estimate of the state vector. The initiation of the Kalman Filter requires the determination of the initial condition, which might be either an assumption or derived from the first available measurement. The Kalman Filter also utilizes the covariance of the estimated error to probabilistically ascertain the updated state estimate, symbolized by the matrix P. The disparity between the true state and the estimated state represents the estimation error, serving as a probabilistic measure of the closeness between the estimated and the true state. In subsequent discussions, the formulation of the equations that the Kalman Filter employs to compute the estimated state and covariance will be elaborated upon meticulously, with each step illustrated using relevant examples to enhance understanding of the subject matter [40][41].

## VI. TEST EXECUTION

In the forthcoming analysis, we embark on a comprehensive exploration of the Kalman Filter, employing a specific example to elucidate the underlying concepts. The focus of our study is the development of a two-dimensional tracking filter aimed at estimating both the position and velocity of a moving object based solely on external measurements. The challenge herein lies in dealing with an uncooperative target, devoid of any onboard sensors to provide direct information such as acceleration. The reliance is entirely on external sensors, which may include radar among other technologies, to gather data about the object's movement.

The application spectrum of such a filter spans numerous domains, including but not limited to, radar-based aircraft tracking, optical tracking for autonomous vehicles to detect and navigate around obstacles or pedestrians, object recognition within image sequences for surveillance and military strategies, robotics, and GPS-enhanced navigation systems. The primary objective is to discern the trajectory of an entity moving within a two-dimensional plane—defined by the x and y axes—utilizing its velocity and deducing its precise location over time. This entails constructing a dynamic model to represent the object's motion, alongside assimilating position measurements derived from external sensing mechanisms.

To address this problem, we introduce a series of Python scripts designed to simulate a two-dimensional tracking scenario. These scripts form the foundation for implementing a Kalman Filter tailored to this context. A deep dive into the code structure and its execution will reveal how the simulation models the filtering process in real-time, encompassing the visualization of the filtering action, innovation from measurements, and a comparative analysis of the true versus estimated states of the system.

The initial graphical output displays the position measurements and their corresponding estimations, highlighting the Kalman Filter's efficacy in tracking the object's true path. Additionally, we scrutinize the innovation process, which signifies the discrepancy between actual and predicted measurements, further illustrated by subsequent plots that detail the positional and velocity states alongside their estimated counterparts.

This endeavor aims to furnish a robust understanding of the Kalman Filter's application in dynamic system tracking through practical exercises. Engaging with the provided Python scripts, readers are guided to incrementally build upon the initial framework, culminating in a fully functional simulation that exemplifies the filter's precision in estimating and tracking the movement of objects within a two-dimensional space. This pedagogical approach ensures a comprehensive grasp of the Kalman Filter's mechanics and its practical utility in real-world scenarios.

### A. Estimation of Pendulum Dynamics

Following our exploration of the linear Kalman Filter, we now proceed to illustrate its applicability to systems with distinct characteristics through an exemplary case study. This discussion centers on the challenge of pendulum state estimation, offering a quintessential scenario to demonstrate the feasibility of employing a linear Kalman Filter for nonlinear systems, given the necessary theoretical adjustments and approximations are integrated within the filter's framework.

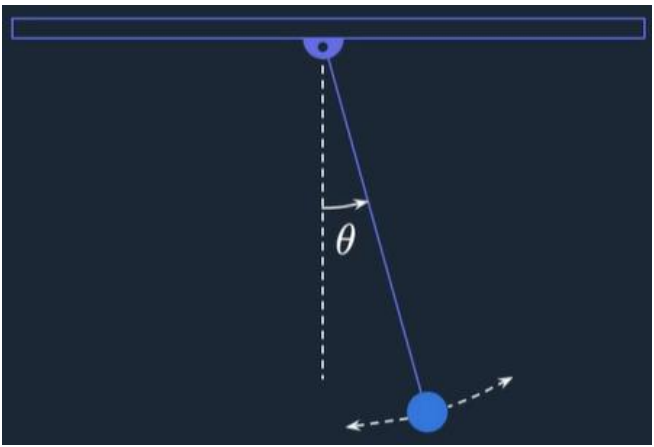

Figure 1: Pendulum Dynamics

$$\text{Estimation:} \quad \hat{x} = [\theta, \dot{\theta}]^T \tag{75}$$

$$\text{Measurement:} \quad z = \theta + v \tag{76}$$

The focal point of this case study is a simple pendulum system, characterized by a mass suspended from a pivot point, which permits unobstructed oscillatory motion. Initiated from an angle deviating from the vertical, the mass undergoes a series of swings. The crux of our estimation problem lies in deducing the pendulum's state—specifically, its angular position ($\theta$) and angular velocity ($\dot{\theta}$)—relying solely on noisy observations of the pendulum's angle ($\theta$).

This investigative journey into pendulum dynamics estimation is structured into several sequential segments, each designed to progressively unravel the complexities of the problem and elucidate the implementation nuances of the Kalman Filter in this context. The initial segment presents an overview of the pendulum estimation challenge, setting the stage for a deeper inquiry into the subject matter. Subsequently, we delve into the dynamics governing the pendulum's motion and the intricacies of the measurement process, laying the groundwork for the filter's application.

Following this, attention shifts to the practical aspects of modeling and executing the Kalman Filter tailored to the pendulum system. This phase is instrumental in bridging theoretical principles with empirical application, highlighting the model's adaptability to the pendulum's dynamics. The penultimate segment focuses on refining the filter's parameters and optimizing its performance, a critical step in ensuring the accuracy and reliability of the state estimates.

The culmination of this examination is a comprehensive summary that encapsulates key insights and learnings derived from applying the Kalman Filter to the pendulum estimation problem. Through this structured approach, we aim to convey the versatility of the Kalman Filter and its potential to address the complexities inherent in nonlinear dynamic systems.

### B. Dynamics of the System and Observations

The pendulum system consists of a mass *m* suspended from a pivot, facilitating its unimpeded oscillation. The essence of our inquiry delves into the decomposition of the system, revealing a mass *m* at one extremity of an ostensibly massless rod of length *L*, the latter anchored at a pivot point. The orientation of this rod is quantified by an angle $\theta$, and its angular velocity by $\dot{\theta}$. The introduction of torques, $\tau$, serves either as an external stimulus to the system or remains null to observe its inherent oscillatory response.

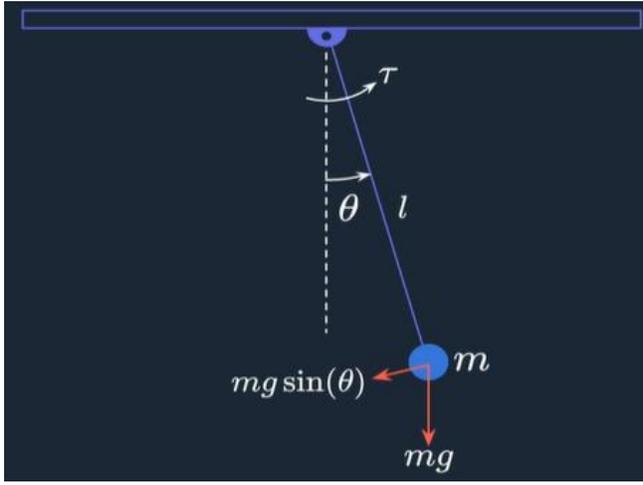

Figure 2: Pendulum System Dynamics

The formulation of the motion's equation, a cornerstone of our analysis, emerges from an integration of the forces and inertial properties governing the pendulum. This culminates in a set of non-linear differential equations adeptly describing the pendulum's dynamic behavior. Notably, the non-linear nature of these equations, encapsulated within an ordinary differential equation framework, necessitates a nuanced approach to modeling.

Pendulum Nonlinear ODE:

$$\ddot{\theta} + \frac{g}{l}\sin(\theta) = \frac{1}{ml^2}\tau \qquad (77)$$

Non-Linear State Space:

$$x_1 = \theta \qquad (78)$$

$$\dot{x}_1 = \dot{\theta} = x_2 \qquad (79)$$

$$\dot{x}_2 = \ddot{\theta} \approx -\frac{g}{l}x_1 + \frac{1}{ml^2}\tau \qquad (80)$$

Linear State Space:

$$\begin{bmatrix}\theta \\ \dot{\theta}\end{bmatrix} = \begin{bmatrix}0 & 1 \\ -\frac{g}{l} & 0\end{bmatrix}\begin{bmatrix}\theta \\ \dot{\theta}\end{bmatrix} + \begin{bmatrix}0 \\ \frac{1}{ml^2}\end{bmatrix}\tau \qquad (81)$$

Commencing with a methodical breakdown of the system, we transition to a representation encompassing two first-order differential systems, thereby delineating the states $X_1$ and $X_2$ as corresponding to $\theta$ and $\dot{\theta}$, respectively. This bifurcation allows for a precise characterization of the system's state dynamics.

Given the premise that the state $X_1$ or $\theta$ remains confined to minor angular deviations—specifically, within a 15-degree threshold—an approximation is employed. This approximation permits the simplification of the sine $\theta$ term to $\theta$ itself, facilitating a linear approximation of the system.

Subsequently, the system is articulated in a linear continuous state-space form, an approximation that replaces the sine term with its linear counterpart, thereby yielding a simplified state-space representation.

This linearized model enables a transformation into a discrete formulation, leveraging the matrix exponential relationship, a technique previously utilized for filtering and estimation applications. This foundation paves the way for constructing a dual-faceted model, encompassing both linear and non-linear approximations, to simulate and contrast the pendulum system's dynamics.

Embedded within this analysis is a Python script, specifically designed for simulating the unforced or free-response behavior of both linear and non-linear pendulum models. A meticulous setup precedes the simulation, establishing parameters such as the time step, overall simulation duration, gravitational constant, pendulum length, and initial angular displacement.

The empirical phase of our investigation reveals pivotal insights. Initially, with the pendulum positioned at a 10-degree displacement, a simulation elucidates the congruence between the linear and non-linear models, manifesting minimal variation in their respective responses. However, adjusting the initial angle to 45 degrees—a value surpassing the linear approximation's validity—unveils significant discrepancies between the models. This divergence accentuates the imperative of adhering to the linear approximation's applicable range to ensure its fidelity.

Ultimately, this detailed exploration within the test execution's framework accentuates the intricacies of modeling dynamic systems. By juxtaposing linear and non-linear models under varied initial conditions, we underscore the criticality of recognizing the limitations inherent to linear approximations. This endeavor not only enriches our understanding of dynamic system behavior but also highlights the importance of methodical model selection and parameterization in achieving accurate simulation outcomes.

C. Implementation of the Kalman Filter Model

Within this segment of our discourse on test execution, attention is drawn to the refinement and application of a linear approximation for the dynamical states estimation within a pendulum system, based on angular measurements denoted as $\theta$. This development phase leverages the linear approximation in conjunction with a linear filter, aiming to enhance the precision of state estimation through the integration of measurement data.

The procedural essence for applying the linear Kalman filter to the stated estimation challenge involves a series of methodical steps. These include the delineation of the state vector $X$, the process model $F$, alongside the articulation of uncertainties through the process uncertainty covariance matrix $Q$, the measurement model $H$, and the measurement uncertainty covariance matrix $R$. These components form the backbone of any estimation task, underpinning the calculations requisite for diverse applications.

Our analytical framework is anchored in a continuous linear system model expressed as $\dot{x} = Ax + Bu$, with the states comprising $\theta$ and $\dot{\theta}$, and an assumption of null external torque, mirroring the approach adopted for analogous 2D tracking problems. This paradigm emphasizes the system's natural or unforced response, necessitating a transition to a discrete model. This transition employs the matrix exponential method, a technique previously elucidated, to encapsulate the dynamics effectively.

Linear Continuous System:

$$\begin{bmatrix} \dot{\theta} \\ \ddot{\theta} \end{bmatrix} = \begin{bmatrix} 0 & 1 \\ -\frac{g}{l} & 0 \end{bmatrix} \begin{bmatrix} \theta \\ \dot{\theta} \end{bmatrix} + \begin{bmatrix} 0 \\ \frac{1}{ml^2} \end{bmatrix} u \quad (82)$$

$$\dot{x} = Ax + Bu \quad (83)$$

An assumption of stochastic, unknown torque, represented by a random variable $w$, introduces a layer of complexity, impacting primarily the $\theta$ state. The variance of this process noise is governed by $\sigma_\tau^2$, which informs the process model noise covariance matrix, thus integrating an additional layer of uncertainty with each predictive step. This strategy aids in accommodating discrepancies arising from imperfect model approximations, enhancing the robustness of the estimation process.

Linear Discrete System:

$$\begin{bmatrix} \theta \\ \dot{\theta} \end{bmatrix}_k = F \begin{bmatrix} \theta \\ \dot{\theta} \end{bmatrix}_{k-1} + \begin{bmatrix} 0 \\ 1 \end{bmatrix} w \quad (84)$$

$$F = e^{A\Delta t} \quad (85)$$

The measurement model adopts a simplistic yet effective approach, capturing angular position measurements with an inherent Gaussian noise component $v$, characterized by zero mean and variance $\sigma_\theta^2$. This setup, coupled with a measurement model that selectively extracts the $\theta$ state, provides a foundation for simulating and implementing the Kalman filter.

Measurement Model:

$$z = \theta + v \quad (86)$$

$$v \sim N(0, \sigma_\theta^2) \quad (87)$$

$$z = \begin{bmatrix} 1 & 0 \end{bmatrix} \begin{bmatrix} \theta \\ \dot{\theta} \end{bmatrix} + v \quad (88)$$

Embarking on a simulation exercise, akin to previous endeavors, participants are tasked with actualizing the outlined equations. The assignment necessitates modifications to the initialization function within the Kalman filter's Python code, specifically updating the $F$, $H$, and associated matrices, to reflect the pendulum system's dynamics accurately.

The simulation scenario is meticulously configured, encapsulating a 10-second timeframe with measurements of $\theta$ at a 10 Hz frequency, each imbued with Gaussian noise. The setup ensures that the initial conditions fall within the linear approximation's valid range, supplemented by animations and plots to visualize the dynamics and the efficacy of the Kalman filter in real-time.

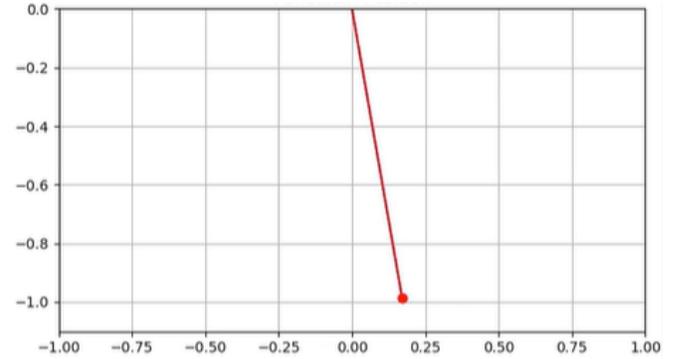

Figure 3: Pendulum Position

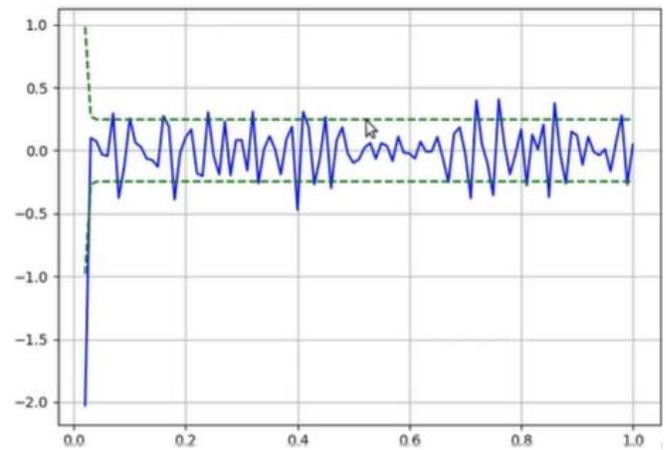

Figure 4: Measurement Innovation

The implementation aims to synchronize the measurement noise characteristics within the simulation with those assumed in the Kalman filter configuration. This coherence is crucial for the filter's initial condition setup, where the first $\theta$ measurement informs the initial state estimate, setting the stage for subsequent predictive and update cycles.

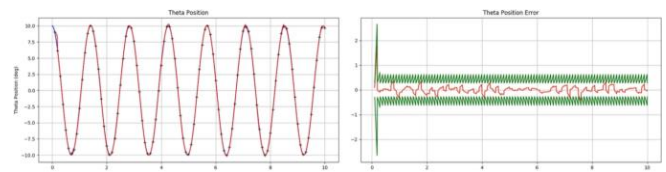

Figure 5: Estimated Theta Angle

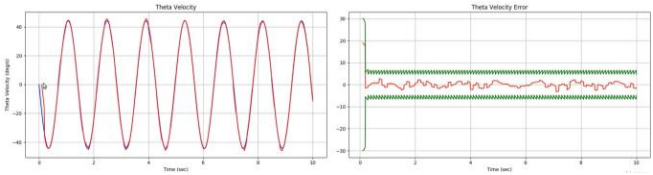

Figure 6: Estimated Theta Velocity

Upon execution, the results affirm the Kalman filter's capability to track the pendulum's true motion closely, with the estimated states and measurement innovations reflecting a high degree of accuracy. The error margins, encapsulated within the three-sigma bounds, attest to the precision of the estimation, underscoring the filter's effectiveness in navigating the inherent uncertainties.

This detailed exploration, integral to the test execution phase, not only underscores the practical application of linear system models and Kalman filtering but also illustrates the nuanced approach required to tailor these methodologies to specific dynamic systems. As we transition to subsequent discussions on filter tuning and performance optimization, the insights garnered here lay a foundational understanding, pivotal for advancing in the domain of dynamic state estimation.

### D. *Optimization and Evaluation of Kalman Filter Performance*

In the current phase of our detailed analysis pertaining to the execution of tests, we delve into a nuanced analysis of a linear filter's application for state estimation in a pendulum system. This examination is predicated on the successful derivation and implementation of a linear approximation model, which serves as a foundational element in our estimation process. The focus herein is on evaluating the filter's performance through empirical analysis and identifying the implications of process noise and model uncertainty on estimation accuracy.

The operational efficacy of the linear filter is initially demonstrated through a simulation, where the pendulum's oscillatory motion is captured, and the estimation accuracy is visually represented. The estimations closely align with the true state of the system, as evidenced by the concurrence of the estimated values (denoted by a red line) with the actual system states (denoted by a blue line). A closer inspection of the measurement innovations reveals a near-zero mean and a variance indicative of a well-calibrated estimation process, with the covariance of the innovation and measurement noise closely matched, underscoring the filter's precision in tracking the pendulum's dynamics.

Further analysis is conducted on the theta position and velocity estimates, which exhibit a high degree of accuracy and fidelity to the system's true states. The errors associated with these estimates are contained within the three-sigma bounds, affirming the filter's robust performance and fine-tuning. The introduction of a nonzero torque standard deviation, a parameter integral to the process noise model, illustrates the critical role of incorporating uncertainty within the system to account for the linear approximation's inherent limitations.

A comparative analysis is undertaken by adjusting the torque standard deviation to zero, thereby eliminating additional process noise from the system. This adjustment yields a discernible impact on the system's behavior, with measurement innovations indicating an incremental divergence over time, suggesting an escalation in the discrepancy between the estimated and actual states. This deviation becomes more pronounced with the extension of the simulation duration, highlighting the filter's diminished efficacy in the absence of adequate process noise to mitigate modeling inaccuracies.

Reinstating the original torque standard deviation and extending the simulation duration further elucidates the necessity of incorporating a measured degree of uncertainty to counterbalance the linear model's simplifications. This approach significantly enhances the filter's adaptability, enabling it to maintain estimation accuracy over prolonged periods and under varying initial conditions.

The exploration extends to scenarios involving larger initial angular displacements, diverging from the small angle assumption. Despite the linear model's theoretical limitations in capturing nonlinear dynamics, the adjusted uncertainty levels within the filter facilitate a commendable degree of accuracy in state estimation, even when confronted with more pronounced nonlinear behaviors. This finding underscores the capacity of a linear filter, equipped with a judiciously calibrated uncertainty model, to effectively approximate nonlinear systems under certain conditions.

This comprehensive analysis demonstrates that, through strategic manipulation of model uncertainty and process noise, a linear filter can be adeptly applied to nonlinear dynamic systems, such as the pendulum model under consideration. The key lies in the systematic adjustment of the uncertainty parameters to compensate for the limitations inherent in linear approximations, thereby ensuring the continued relevance and applicability of linear estimation methodologies in complex, nonlinear contexts. This insight not only validates the robustness of the linear filter in a broad array of scenarios but also highlights the importance of nuanced model calibration in achieving optimal estimation performance.

### VII. SUMMARY

In this concluding section of our research, we delve into the intricate process of applying filtering techniques to nonlinear dynamical systems, a topic that bridges theoretical modeling with practical application. The essence of our investigation centers on the viability of employing filters, traditionally designed for linear systems, within the context of nonlinear system analysis. This endeavor necessitates a nuanced approach, wherein appropriate assumptions or approximations are strategically deployed to reformulate the nonlinear system characteristics into a linear paradigm. Such a transformation, while facilitating the use of linear filters, introduces complexities in the modeling process, notably an increase in error.

A pivotal aspect of our discourse is the acknowledgment of the inherent trade-offs involved in augmenting the process model noise to counterbalance the increased error resulting from linear approximations. This adjustment, essential for maintaining the filter's operational integrity, inevitably impacts the filter's reliance on its predictive model versus actual measurements. Specifically, an elevation in process noise diminishes the weight of the prediction model, thereby compelling the filter to lean more heavily on real-time measurements for state estimation. This shift, while necessary, affects the filter's performance, necessitating a delicate balance to optimize its functionality.

Further complicating the application of linear filters to nonlinear systems is the necessity for meticulous testing and manual tuning. This process is critical for ensuring optimal filter performance, underscoring the importance of a hands-on approach to filter configuration. The calibration of the filter, particularly in adjusting for increased process model noise, is paramount to its success in accurately capturing the dynamics of the system it seeks to model.

One of the more challenging aspects of applying linear filters to nonlinear systems is the potential for divergence from the true state. This divergence is attributed to the accumulation of errors stemming from the numerical approximations employed in the modeling process. As the model deviates from the real-world system, it aims to represent, these discrepancies can compound over time, leading to a significant departure from the actual system states. The risk of divergence underscores the criticality of precision in the modeling phase and the judicious adjustment of process noise to mitigate the effects of approximation errors.

To augment our analysis, we explored various scenarios wherein the application of linear filters to nonlinear systems was both successful and challenging. These case studies provided valuable insights into the factors that contribute to the effective adaptation of linear filters for nonlinear contexts. Through rigorous simulation and empirical validation, we demonstrated the feasibility of this approach, while also highlighting the limitations and considerations that must be addressed to achieve accurate state estimation.

In summary, our research contributes to the broader understanding of how linear filters can be adapted for nonlinear systems, provided that careful adjustments and calibrations are made to accommodate the nuances of nonlinear dynamics. The study bridges theoretical concepts with practical applications, offering a roadmap for researchers and practitioners alike to navigate the complexities of applying linear filtering techniques to nonlinear systems. Through a combination of theoretical exploration and empirical analysis, we have illuminated the path forward in this challenging yet rewarding domain, marking a significant step in the evolution of dynamic system modeling and analysis.

## VIII. Conclusion

As our comprehensive exploration of the linear Kalman filter and its application in state estimation and sensor fusion draws to a close, we seize this moment to reflect on the journey undertaken and the breadth of knowledge acquired. Our expedition through the realms of probabilistic modeling, system dynamics, and optimal estimation techniques has equipped us with a profound understanding of the theoretical underpinnings and practical applications of the linear Kalman filter.

Beginning with the fundamentals, we delved into the probabilistic representation of uncertainty, utilizing the Gaussian distribution as a cornerstone for estimating the state and measurement uncertainties. This foundational knowledge set the stage for our exploration of converting differential equations into state-space representations, a critical step for both linear and nonlinear systems. The capacity to simulate these dynamic systems underpins the practical application of the theories discussed, bridging the gap between abstract concepts and their real-world implementations.

Our inquiry extended to the realm of estimation techniques, where we explored the least squares estimation method as a solution for static estimation challenges. Building upon this, we ventured into the more complex territory of the linear Kalman filter, dedicating our focus to solving optimal estimation problems. This segment not only constituted the core of our investigation but also highlighted the versatility and efficacy of the Kalman filter in a wide array of applications.

The journey did not stop at theoretical exploration; we also embarked on the practical application of these concepts through the implementation of the linear Kalman filter in Python. This hands-on experience underscored the universality of the mathematical principles underpinning the Kalman filter, demonstrating their applicability across different programming languages and real-world scenarios.

In synthesizing the knowledge acquired, we now possess a robust framework for approaching sensor fusion and state estimation challenges. This framework is bolstered by an arsenal of resources, including reference documents and cheat sheets, designed to support and guide future endeavors in implementing custom solutions or navigating complex estimation problems.

In closing, this journey through the intricacies of the linear Kalman filter was crafted to empower you with the skills and insights necessary to become a subject matter expert in the field. The comprehensive coverage of both theory and application aims to leave you well-prepared to tackle a diverse array of problems in sensor fusion and state estimation. It is with a sense of accomplishment and anticipation of your future successes that we extend our heartfelt thanks for engaging with this material. Your dedication to mastering the content presented herein not only enriches your knowledge base but also positions you for continued growth and exploration in this dynamic field. As you move forward, may the principles and techniques explored serve as both a foundation and a springboard for your continued learning and professional development in the realm of Kalman filtering and beyond.